\RequirePackage{ifthen}

\newcommand{\typeof}{0} %
\newcommand{\cmnts}{0} %
\newcommand{\longv}[1]{\ifthenelse{\equal{\typeof}{0}}{}{#1}}
\newcommand{\shortv}[1]{\ifthenelse{\equal{\typeof}{0}}{#1}{}}
\newcommand{\longshortv}[2]{\ifthenelse{\equal{\typeof}{0}}{#2}{#1}}
\newcommand{\nocommentsv}[1]{\ifthenelse{\equal{\cmnts}{0}}{#1}{}}
\newcommand{\commentsv}[1]{\ifthenelse{\equal{\cmnts}{0}}{}{#1}}

\documentclass[submission]{eptcs}
\providecommand{\event}{DICE-FOPARA 2017}
\usepackage{underscore}

\usepackage{microtype}
\usepackage{enumerate}
\usepackage{textcomp}
\usepackage{latexsym} 
\usepackage{amsfonts} 
\usepackage{amssymb}
\usepackage{amsmath}
\usepackage{hyperref}
\usepackage{pgf}
\usepackage{tikz}
\usepackage{subcaption}
\usepackage{cleveref}

\usepackage{general}
\usepackage{sizetypes}

\title{Automating Sized-Type Inference\\ and Complexity Analysis}
\def\titlerunning{Automating Sized-Type Inference and Complexity Analysis}
\def\authorrunning{M. Avanzini \& U. Dal Lago}
\date{}
\author{Martin Avanzini \and Ugo Dal Lago}

\renewcommand{\labelitemi}{--}

\makeatletter
\renewcommand{\paragraph}{%
  \@startsection{paragraph}{4}%
  {\z@}{1ex \@plus 1ex \@minus .2ex}{-1em}%
  {\normalfont\normalsize\em}%
}
\makeatother

\newtheorem{theorem}{Theorem}

\newtheorem{definition}{Definition}

\newtheorem{corollary}{Corollary}

\newcounter{commentcounter}

\begin{document}

\maketitle

\begin{abstract}
This paper introduces a new methodology for the complexity analysis of
higher-order functional programs, which is based on three components:
a powerful type system for size analysis and a sound type inference
procedure for it, a ticking monadic transformation and a concrete tool
for constraint solving. Noticeably, the presented methodology can be
fully automated, and is able to analyse a series of examples which
cannot be handled by most competitor methodologies. This is possible
due to various key ingredients, and in particular an abstract index
language and index polymorphism at higher ranks. A prototype
implementation is available.
\end{abstract}

%%%%%%%%%%%%%%%%%%%%%%
\section{Introduction}
%%%%%%%%%%%%%%%%%%%%%%

One successful approach to automatic verification of termination
properties of higher-order functional programs is based on \emph{sized
  types}~\cite{HPS:POPL:96}. In sized types,
a type carries not only some information about the \emph{kind} of each
object, but also about its \emph{size}, hence the name.  This
information is then exploited when requiring that recursive calls are
done on arguments of \emph{strictly smaller} size.  Estimating the
size of intermediate results is crucial for complexity
analysis, but up to now, the only attempt of using sized types for
complexity analysis is due to Vasconcelos~\cite{Vasconcelos:Diss:08},
and confined to space complexity. If one wants to be sound for time
analysis, size types need to be further refined, e.g., by turning
them into linear dependently types~\cite{LG:LMCS:11}. % correct cite?

%MA: new
Since the first inception in the seminal paper of Hughes~et.~al.~\cite{HPS:POPL:96}
the literature on sized typed has grown to a considerable extend. 
Indeed, various significantly more expressive systems have been introduced, 
with the main aim to improve the expressiveness in the context of termination analysis. 
For instance, Blanqui~\cite{Blanqui:CSL:05} introduced a novel sized type system 
on top of the \emph{calculus of algebraic construction}. Notably, it has been 
shown that for size indices over the successor algebra, type checking is decidable~\cite{Blanqui:CSL:05}. 
The system is thus capable of expressing additive relations between sizes. In 
the context of termination analysis, where one would like to statically detect that 
a recursion parameter decreases in size, this is sufficient. 
In this line of research falls also more recent work of Abel and Pientka~\cite{AP:JFP:16}, 
where a novel sized type system for termination analysis on top of $\mathsf{F}_\omega$ is proposed.
Noteworthy, this system has been integrated in the dependently typed language~\tool{Agda}.\footnote{See \url{http://wiki.portal.chalmers.se/agda}.}
%

%MA: intro changed
As we will see, capturing only additive relations between value sizes 
is not enough for our purpose. On the other hand, even slight extensions to the size index language
render current methods for type inference, even type checking, intractable. 
In this paper, we thus take a fresh look at sized-type systems, with a particular 
emphasis on a richer index language and feasible automation on existing constraint solving technology. 
Our system exhibits many similarities with the archetypal system from~\cite{HPS:POPL:96}, 
which itself is based on a Hindley-Milner style system. 
Although conceptually simple, our system is substantially more expressive than the
traditional one. This is possible mainly due to the addition of one ingredient, 
viz, the presence of \emph{arbitrary rank index polymorphism}. That is, functions that take
functions as their argument can be polymorphic in their size
annotation.  
Of course, our sized type system is proven a sound methodology for \emph{size} analysis. 
In contrast to existing works, one can also device an inference machinery that 
is sound and (relative) complete. Finally, this system
system is amenable to time complexity analysis by a
ticking monadic transformation. A prototype implementation is
available, see below for more details. More specifically, our
contributions can be summarized as follows:
\begin{varitemize}
\item
  We show that size types can be generalised so as to encompass a
  notion of index polymorphism, in which (higher-order subtypes of)
  the underlying type can be universally quantified. This allows for a
  more flexible treatment of higher-order functions. Noticeably, this
  is shown to preserve soundness (i.e. subject reduction), the minimal
  property one expects from such a type system. On the one hand, this
  is enough to be sure that types reflect the size of the underlying
  program. On the other hand, termination is not enforced anymore by
  the type system, contrarily to, e.g.~\cite{Blanqui:CSL:05,AP:JFP:16}. In
  particular, we do not require that recursive calls are made on
  arguments of smaller size. 
% \item
%   The type inference problem is shown to be (relatively) decidable by
%   giving an algorithm which, given a program, produces in output
%   candidate types for the program, together with a set of integer index
%   inequalities which need to be checked for satisfiability. This style
%   of results is quite common in sized types. 
%   %MA: added
%   In contrast to existing works, we put only mild assumptions on the index language, 
%   and we do not require that the generated inequalities admit a most general solution. 
%   Indeed, this enables us to express significantly more complicated size relations 
%   between inputs and outputs. 
\item
  The polymorphic sized types system, by itself, does not guarantee
  any complexity-theoretic property on the typed program, except for
  the \emph{size} of the output being bounded by a function on the
  size of the input, itself readable from the type. Complexity
  analysis of a program $\progone$ can however be seen as a size analysis of
  another program $\tprogone$ which computes not only $\progone$, but its
  complexity. This transformation, called the \emph{ticking
    transformation}, has already been studied in similar
  settings~\cite{DLR:ICFP:15}.
\item
  Contrarily to many papers from the literature, we have taken care
  not only of constraint \emph{inference}, but also of constraint
  \emph{solving}.  
  %MA: changed
  This has been done by building a prototype called \tool{HoSA} which
  implements type inference and ticking, and then relies on an external tool, 
  called \tool{GUBS}, to check the generated constraints for satisfiability.  
  %MA: extended
  \tool{GUBS} borrows heavily from the advances made over the last decade 
  in the synthesis of \emph{polynomial interpretations}, a form of polynomial ranking function, 
  by the rewriting community. It features also some novel aspects, most importantly, 
  a bottom-up SCC analysis for incremental constraint solving. 
  We thus arrive at a fully automated runtime analysis of higher-order functional programs.
  Noteworthy, we are able to effectively infer polynomial, not necessarily linear, bounds on the 
  runtime of programs. 
  
  Both tools are open source and available from the first authors homepage.%
  \footnote{See \url{https://cl-informatik.uibk.ac.at/users/zini/software}.}
  \tool{HoSA}\ is able to analyse, fully automatically, a series of examples which
  cannot be handled by most competitor methodologies. 
  Indeed, it is to our best knowledge up until today the only approach
  that can fully deal with function closures whose complexity depends on the captured environment,
  compare for instance the very recent work of Hoffmann~et.~al. \cite{HDW:POPL:17}.
  Dealing with such closures is of crucial importance, e.g., 
  when passing partially applied functions to higher-order combinators, a 
  feature pervasively used in functional programming. 
\end{varitemize}

For brevity, we only give a formalisation of our system and state the 
central theorem here. An extended version, including 
all the technical details is available online~\cite{EV}.
%%%%%%%%%%%%%%%%%%%%%%%%%%%%%%%%%%%%%
\section{Our Type System at a Glance}\label{sect:ERW}
%%%%%%%%%%%%%%%%%%%%%%%%%%%%%%%%%%%%%
In this section, we will motivate the design choices we made when
defining our type system through some examples. This can also be taken
as a gentle introduction to the system for those readers which are
familiar with functional programming and type theory. Our type system
shares quite some similarities with the prototypical system introduces
by Hughes~et.~al.~\cite{HPS:POPL:96} and similar ones \cite{GBR:CSL:08,Vasconcelos:Diss:08}, 
but we try to keep presentation as self-contained as possible.
%%%%%%%%%%%%%%%%%%%
\paragraph{Basics.}
%%%%%%%%%%%%%%%%%%%
We work with functional programs over a fixed set of inductive
datatypes, e.g. $\tynat$ for natural numbers and $\tylist{\alpha}$ for
lists over elements of type $\tyvarone$.  Each such datatype is
associated with a set of typed \emph{constructors}, below we will use
the constructors $0 \decl \tynat{}$, $\con{Succ} \decl \tynat \starr
\tynat$ for naturals, and the constructors $\cnil \decl \forall
\tyvarone.\ \tylist{\tyvarone}$ and the infix constructor $(\ccons)
\decl \forall \tyvarone.\ \tyvarone \starr \tylist{\tyvarone} \starr
\tylist{\tyvarone}$ for lists.  Sized types refine each such datatype
into a \emph{family} of datatypes indexed by natural
numbers, their \emph{size}. E.g., to $\tynat$ and
%MA: changed: constants get size 0
$\tylist{\tyvarone}$ we associate the families
$\tynat[0],\tynat[1],\tynat[2],\dots$ and
$\tylist[0]{\tyvarone},\tylist[1]{\tyvarone},\tylist[2]{\tyvarone},\dots$,
respectively. 
An indexed datatype such as $\tylist[n]{\tynat[m]}$
then represents lists of length $n$, over naturals of size $m$. 

A function $\fun{f}$ will then be given a polymorphic type $\forall
\vec{\tyvarone}.\ \forall\vec{\ivarone}.\ \mtpone \to \mtptwo$. 
Whereas the variables $\vec{\tyvarone}$ range over types, the variables
$\vec{\ivarone}$ range over sizes.  Datatypes
occurring in the types $\mtpone$ and $\mtptwo$ will be indexed by
expressions over the variables $\vec{\ivarone}$. E.g., the append
function can be attributed the sized type $\forall \tyvarone.\ \forall
\ivarone\ivartwo.\ \tylist[\ivarone]{\tyvarone} \starr
\tylist[\ivartwo]{\tyvarone} \starr \tylist[\ivarone +
  \ivartwo]{\tyvarone}$.

Soundness of our type-system will guarantee that when append is
applied to lists of length $n$ and $m$ respectively, will yield a list
of size $n + m$, or possibly diverge. In particular, our type system is
not meant to guarantee termination, and complexity analysis will be done
via the aforementioned ticking transformation, to be described later. As
customary in sized types, we will also integrate a
subtyping relation $\mtpone \subtypeof[] \mtptwo$ into our system,
allowing us to relax size annotations to less precise ones.
% MA: mention higher-order functions
This flexibility is necessary to treat conditionals where the
branches are attributed different sizes, or, to treat higher-order
combinators which are used in multiple contexts.

Our type system, compared to those from the literature, has its main
novelty in polymorphism, but is also different in some key aspects,
addressing intensionality but also practical considerations towards
type inference. In the following, we shortly discuss the main differences.
%%%%%%%%%%%%%%%%%%%%%%%%%%%%%%%%%%%%%%%%
\paragraph{Canonical Polymorphic Types.}
%%%%%%%%%%%%%%%%%%%%%%%%%%%%%%%%%%%%%%%%
We allow polymorphism over size expressions, but put some syntactic
restrictions on function declarations: In essence, we disallow
non-variable size annotations directly to the left of an arrow, and
furthermore, all these variables must be pairwise distinct.  We call
such types canonical. The first restriction dictates that
e.g. $\fun{half} \decl \forall \ivarone. \tynat[2 \cdot i] \starr
\tynat[i]$ has to be written as $\fun{half} \decl \forall
\ivarone. \tynat[i] \starr \tynat[i/2]$.  The second restriction
prohibits e.g.\ the type declaration $\fun{f} \decl \forall
\ivarone. \tynat[\ivarone] \starr \tynat[\ivarone] \starr \mtpone$,
rather, we have to declare $\fun{f}$ with a more general type 
$\forall \ivarone \ivartwo. \tynat[\ivarone] \starr \tynat[\ivartwo] \starr \mtpone'$.  
The two restrictions considerably simplify the inference
machinery when dealing with pattern matching, and pave the way towards
automation.  Instead of a complicated unification based mechanism, a
matching mechanism suffices. 
\paragraph{Abstract Index Language.}
%%%%%%%%%%%%%%%%%%%%%%%%%%%%%%%%%%%%
Unlike in~\cite{HPS:POPL:96}, where indices are formed over naturals
and addition, we keep the index language abstract. This allows for
more flexibility, and ultimately for a better intensionality. Indeed,
having the freedom of not adopting a fixed index language is known
to lead towards completeness~\cite{LG:LMCS:11}.
%%%%%%%%%%%%%%%%%%%%%%%%%%%%%%%%%%%%%%%%%%%%%
\paragraph{Polymorphic Recursion over Sizes.}
%%%%%%%%%%%%%%%%%%%%%%%%%%%%%%%%%%%%%%%%%%%%% 
\begin{figure}[t]
  \begin{framed}
    \begin{minipage}{1.0\linewidth}
\begin{lstlisting}[emph={i,j,k,l,ijk,ij,x,xs,f,ys},emph={[2] reverse, rev},style=haskell, style=numbered]
rev $\decl \forall \tyvarone.\ \forall \ivarone\ivartwo.\ \tylist[\ivarone]{\tyvarone} \starr \tylist[\ivartwo]{\tyvarone} \starr \tylist[\ivarone + \ivartwo]{\tyvarone}$
rev []       ys = ys
rev (x : xs) ys = rev xs (x : ys)

reverse $\decl \forall \tyvarone.\ \forall \ivarone.\ \tylist[\ivarone]{\tyvarone} \starr \tylist[\ivarone]{\tyvarone}$
reverse xs = rev xs [] 
\end{lstlisting}
    \end{minipage}
  \end{framed}
  \caption{Sized type annotated tail-recursive list reversal function.}\label{fig:reverse}
\end{figure}
Type inference in functional programming languages, such as \haskell\ or \ocaml, 
is restricted to parametric polymorphism in the form of \emph{let-polymorphism}.
Recursive definitions are checked under a monotype, thus, 
types cannot change between recursive calls. 
Recursive functions that require full parametric polymorphism~\cite{Mycroft:ISP:84} have to be 
annotated in general, as type inference is undecidable in this setting.

Let-polymorphism poses a significant restriction in our context, because sized types considerably refine upon simple types.
Consider for instance the usual tail-recursive definition of list reversal depicted in Figure~\ref{fig:reverse}.
With respect to the annotated sized types, 
in the body of the auxiliary function $\fun{rev}$ defined on line 4, 
the type of the second argument to $\fun{rev}$ will change from $\tylist[\ivartwo]{\tyvarone}$ (the assumed type of $ys$) to
$\tylist[\ivartwo+1]{\tyvarone}$ (the inferred type of $x \ccons ys$). Consequently, $\fun{rev}$ is not typeable under a monomorphic sized type.
Thus, to handle even such very simple functions, we will have to overcome let-polymorphism, on the layer of size annotations.
To this end, conceptually we allow also recursive calls to be given a type polymorphic over size variables. 
This is more general than the typing rule for recursive definitions found in more traditional systems~\cite{HPS:POPL:96,GBR:CSL:08}.
%%%%%%%%%%%%%%%%%%%%%%%%%%%%%%%%%%%%%%%%%%%%%%%%%%
\paragraph{Higher-ranked Polymorphism over Sizes.}
%%%%%%%%%%%%%%%%%%%%%%%%%%%%%%%%%%%%%%%%%%%%%%%%%%
In order to remain decidable, classical type inference systems 
work on polymorphic types in \emph{prenex form} $\forall{\vec{\tyvarone}}.\mtpone$, 
where $\mtpone$ is quantifier free.  In our context, it is often not enough to give a
combinator a type in prenex form, in particular when the combinator
uses a functional argument more than once. All uses of the functional
argument have to be given then \emph{the same} type. In the context of sized
types, this means that functional arguments can be applied only to
expressions whose attributed size equals.  
\shortv{%
This happens for instance in recursive combinators, but also non-recursive
ones such as the following function $\fun{twice}  \api f \api x = f \api (f \api x)$.
}\longv{%
This happens for instance
in recursive combinators, but also non-recursive
ones such as the following function $\fun{twice}$.
\begin{align*}
  & \fun{twice} \decl \forall \tyvarone.\ (\tyvarone \starr \tyvarone) \starr \tyvarone \starr \tyvarone\\
  & \fun{twice} \api f \api x = f \api (f \api x)
    \tpkt
\end{align*}}
A strong type-system would allow us to type the expression
$\fun{twice} \api \con{Succ}$ with a sized type $\tynat[c] \starr
\tynat[c + 2]$.  A (specialised) type in prenex form for $\fun{twice}$, such as
\[
  \fun{twice} \decl
  \forall \ivarone.\ (\tynat[\ivarone] \starr \tynat[\ivarone + 1]) \starr \tynat[\ivarone] \starr \tynat[\ivarone + 2]
  \tkom
\]
would immediately yield the mentioned sized type for $\fun{twice} \api
\con{Succ}$.  However, we will not be able to type $\fun{twice}$
itself, because the outer occurrence of $f$ would need to be typed as
$\tynat[\ivarone+1] \starr \tynat[\ivarone + 2]$, whereas the type of
$\fun{twice}$ dictates that $f$ has type $\tynat[\ivarone] \starr
\tynat[\ivarone + 1]$.

The way out is to allow polymorphic types of rank \emph{higher than}
one when it comes to size variables, i.e.\ to allow quantification of
size variables to the left of an arrow at arbitrary depth.  Thus, we can declare
\[
  \fun{twice} \decl
  \forall \ivarone.\ (\forall \ivartwo.\tynat[\ivartwo] \starr \tynat[\ivartwo + 1]) \starr \tynat[\ivarone] \starr \tynat[\ivarone + 2]
  \tpkt
\]
As above, this allows us to type the expression $\fun{twice} \api
\con{Succ}$ as desired.  Moreover, the inner quantifier permits the
two occurrences of the variable $f$ in the body of $\fun{twice}$ to
take types $\tynat[\ivarone] \starr \tynat[\ivarone + 1]$ and
$\tynat[\ivarone+1] \starr \tynat[\ivarone + 2]$ respectively, and
thus $\fun{twice}$ is well-typed.

%%%%%%%%%%%%%%%%%%%%%%%%%%%%%%%%%%%%%%
\paragraph{The Ticking Transformation.}
%%%%%%%%%%%%%%%%%%%%%%%%%%%%%%%%%%%%%%
Our type system only reflect upon one extensional properties
of programs, namely how the size of the output relates to the size of the input.
Runtime analysis can however be reduced to size analysis, e.g. via a \emph{ticking transformation}.
This transformation takes a program $\progone$ 
and translates it into another program $\tprogone$. The transformed program
behaves like $\progone$, 
but additionally computes also the runtime on the given input. 
Technically, the latter is achieved by threading through the computation a counter, 
the \emph{clock}, which is advanced whenever an equation of $\progone$ fires. 
A $k$-ary function 
$\funone \decl \stone_1 \starr \cdots \starr \stone_{k} \starr \stone$
of $\progone$ will be modeled in $\tprogone$ by a function
$\tfun{\funone}{k} \decl \ttype{\stone_1} \starr \cdots \starr \ttype{\stone_{k}} \starr \tclock \starr \tpair{\ttype{\stone}}{\tclock}$,
where $\tclock$ is the type of the \emph{clock}. 
Here, $\ttype{\sttwo}$ enriches functional types $\sttwo$ with clocks accordingly. 
The function $\tfun{\funone}{k}$ behaves in essence like $\funone$, 
but advances the threaded clock suitably.
The clock-type $\tclock$ encodes the running time in unary notation. The size
of the clock thus corresponds to its value.  
Our type system can estimate the size of the clock in the ticked program $\tprogone$, and thus
the runtime of the considered program $\progone$. Noteworthy, this transformation 
is straight forward to implement. 
%%%%%%%%%%%%%%%%%%%%%%%%%%%%%%%
\section{A Worked Out Example}
%%%%%%%%%%%%%%%%%%%%%%%%%%%%%%%
\begin{figure}[t]
  \centering
  \begin{framed}
    \begin{minipage}{1.0\linewidth}
\begin{lstlisting}[emph={f,x,xs,b,ms,ns,ps,m,n},emph={[2] foldr,product},style=haskell, style=numbered]
foldr $\decl \forall \tyvarone\tyvartwo.\ \forall \ivartwo\ivarthree\ivarfour.\ (\forall \ivarone.\ \tyvarone \starr \tylist[\ivarone]{\tyvartwo} \starr \tylist[\ivarone + \ivartwo]{\tyvartwo})
        \starr \tylist[\ivarthree]{\tyvartwo} \starr \tylist[\ivarfour]{\tyvarone} \starr \tylist[\ivarfour \cdot \ivartwo + \ivarthree]{\tyvartwo}$
foldr f b []       = b
foldr f b (x : xs) = f x (foldr f b xs)%\label{fig:doublefilter:foldrec}%

product $\decl \forall \tyvarone \tyvartwo.\ \forall \ivarone\ivartwo.\ \tylist[\ivarone]{\tyvarone} \starr \tylist[\ivartwo]{\tyvartwo} \starr \tylist[\ivarone \cdot \ivartwo]{(\tpair{\tyvarone}{\tyvartwo})}$
product ms ns = foldr (\ m ps. foldr (\ n. (:) (m,n)) ps ns) [] ms
\end{lstlisting}
    \end{minipage}
  \end{framed}
  \caption{Sized type annotated program computing the cross-product of two lists.}\label{fig:doublefilter}
\end{figure}

In this section we give a nontrivial example.
The sized type
annotated program is given in Figure~\ref{fig:doublefilter}.  The
function $\fun{product}$ computes the cross-product $[\,(m,n) \mid m
  \in ms, n \in ns\,]$ for two given lists $ms$ and $ns$. It is
defined in terms of two folds. The inner fold appends, for a fixed
element $m$, the list $[\,(m,n) \mid n \in ns \,]$ to an accumulator
$ps$, the outer fold traverses this function over all elements $m$
from $ms$.
%MA: see hoca
% This, by the way, is an example, that cannot be managed by any of the
% automatic tools from the literature.\footnote{Except our own tool \tool{HoCA}.}%MA: bash a bit our own tool here ;)

In a nutshell, checking that a function $\fun{f}$ is typed correctly
amounts to checking that all its defining equations are well-typed,
i.e.\ under the assumption that the variables are typed according to
the type declaration of $\fun{f}$, the right-hand side of the equation
has to be given the corresponding return-type. Of course, all of this
has to take pattern matching into account.

Let us illustrate this on the recursive equation of $\fun{foldr}$ given in Line~\ref{fig:doublefilter:foldrec} in Figure~\ref{fig:doublefilter}.
Throughout the following, we denote by $\termone \oftype \mtpone$ that
the term $\termone$ has type $\mtpone$.  
To show that the equation is well-typed, let us assume the following 
types for arguments: 
$f \oftype \forall \ivarone.\ \tyvarone \starr \tylist[\ivarone]{\tyvartwo} \starr \tylist[\ivarone + \ivartwo]{\tyvartwo}$, 
$b \oftype \tylist[\ivarthree]{\tyvartwo}$, 
$x \oftype \tyvarone$ and 
$xs \oftype \tylist[m]{\tyvarone}$ for arbitrary size-indices
$\ivartwo,\ivarthree,m$.
Under these assumptions, the left-hand side has type 
$\tylist[(m + 1) \cdot \ivartwo + \ivarthree]{\tyvartwo}$, taking 
into account that the recursion parameter $x \ccons xs$ has size $m+1$.
To show that the equation is well-typed, we verify that the right-hand 
side can be attributed the same sized type.
To this end, we proceed inside out as follows.
\begin{varenumerate}
\item 
  We instantiate the polymorphic type of $\fun{foldr}$ and derive
  \[
    \fun{foldr} \oftype (\forall \ivarone.\ \tyvarone \starr \tylist[\ivarone]{\tyvartwo} \starr \tylist[\ivarone + \ivartwo]{\tyvartwo})
    \starr \tylist[\ivarthree]{\tyvartwo} \starr \tylist[m]{\tyvarone} \starr \tylist[m \cdot \ivartwo + \ivarthree]{\tyvartwo}\tspkt
  \]
\item 
  from this and the above assumptions we get $\fun{foldr} \api f \api b \api xs \oftype
  \tylist[m \cdot \ivartwo + \ivarthree]{\tyvartwo}$;
\item 
  by instantiating the quantified size variable $\ivarone$ in the assumed type of $f$ with 
  the index term $m \cdot \ivartwo + \ivarthree$ we get
  $f \oftype \tyvarone \starr \tylist[m \cdot \ivartwo + \ivarthree]{\tyvartwo} \starr \tylist[(m \cdot \ivartwo + \ivarthree) + \ivartwo]{\tyvartwo}$;
\item 
  from the last two steps we finally get $f \api x \api (\fun{foldr} \api f \api b \api xs) \oftype \tylist[(m + 1) \cdot \ivartwo + \ivarthree]{\tyvartwo}$.
\end{varenumerate}

We will not explain the type checking of the remaining
equations.
However, we would like to stress two crucial points
concerning the type of $\fun{foldr}$.  First of all, we could only
suitably type the two occurrences of $f$ in the body of $\fun{foldr}$
since $f$ was given a type polymorphic in
the size of its arguments.  Secondly, notice that the variable
$\ivartwo$ in the type of $\fun{foldr}$ relates the size of the result
of the argument function to the size of the result of $\fun{foldr}$.
This turns out to be a very useful feature in our system, as any
expression that can be given a type of the form $\typeone \starr
\tylist[k]{\typetwo} \starr \tylist[k+m]{\typetwo}$ is applicable to
$\fun{foldr}$, even if $m$ depends on the environment of the
call-site.  In particular, we will be able to instantiate both
$\lambda$-abstractions in the definition of $\fun{product}$ to such a
type, despite that for the outer abstraction, $m$ depends on the size
of the captured variable $ns$.
It is also worthy of note that the example is only typable since 
the most general type for $\fun{foldr}$, namely 
$(\tyvarone \starr \tyvartwo \starr \tyvartwo) \starr \tyvartwo \starr \tylist{\tyvarone} \starr \tyvartwo$ has been sufficiently instantiated. 
Our implementation \tool{HoSA}\ performs such an instantiation when required, 
and can infer the sized type of $\fun{product}$ specified above automatically. 

%%% Local Variables:
%%% mode: latex
%%% TeX-master: "paper"
%%% End:

%%%%%%%%%%%%%%%%%%%%%%%%%%%%%%%%%%%%%%%%%%%%%%%
\section{Applicative Programs and Simple Types}\label{sect:APST}
%%%%%%%%%%%%%%%%%%%%%%%%%%%%%%%%%%%%%%%%%%%%%%%
We restrict our attention to a small prototypical, strongly typed
functional programming language. For the sake of simplifying
presentation, we impose a simple, monomorphic, type system on
programs, which does not guarantee anything except a form of type
soundness.  We will only later in this paper introduce sized types
proper.  Our theory can be extended straightforwardly to an ML-style
polymorphic type setting. Here, such an extension would only distract
from the essentials. Indeed, our implementation allows polymorphic function definitions.

Let $\STBASE$ denote a finite set of base types $\stbaseone,\stbasetwo, \dots$\,. 
\emph{Simple types} are inductively generated from $\stbaseone\in\STBASE$:
%MA: space
% \begin{align*}
%   & \text{\textbf{(simple types)}} & \stone,\sttwo,\stthree \bnfdef {} 
%   & \stbase && \textit{base type} \\[-1mm]
%   &&\mid\ & \tpair{\stone}{\sttwo} && \textit{pair type} \\[-1mm]
%   &&\mid\ & \stone \starr \sttwo && \textit{function type.}
% \end{align*}
\begin{align*}
  & \text{\textbf{(simple types)}} & \stone,\sttwo,\stthree \bnfdef {} 
  & \stbase \mid\ \stone \starr \sttwo \tpkt
\end{align*}

We follow the usual convention that $\starr$ associates to the right.
Let $\VAR$ denote a countably infinite set of \emph{variables}, ranged
over by metavariables like $\varone$, $\vartwo$.  Furthermore, let
$\FUN$ and $\CON$ denote two disjoint sets of symbols, the set of
\emph{functions} and \emph{constructors}, respectively, all pairwise
distinct with elements from $\VAR$. Functions and constructors are
denoted in \texttt{teletype font}.  We keep the convention that
functions start with a lower-case letter, whereas constructors start
with an upper-case letter.  Each symbol $s \in \VAR \cup \FUN \cup
\CON$ has a simple type $\stone$, and when we want to insist on that,
we write $s^\stone$ instead of just $s$.  Furthermore, each symbol
$s^{\stone_1 \starr \cdots \starr \stone_n \starr \sttwo} \in \FUN
\cup \CON$ is associated a natural number $\arity{s} \leq n$, its
\emph{arity}.  The set of \emph{terms}, \emph{patterns} and \emph{values}
over functions $\funone \in \FUN$, constructors
$\conone \in \CON$ and variables $\varone \in \VAR$ is inductively
generated as follows.  Here, each term receives implicitly a type, in
Church style. Below, we employ the usual convention that application
associates to the left.
\begin{align*}
  & \text{\textbf{(terms)}} & \termone,\termtwo \bnfdef {} 
  & \varone^\stone \mid \funone^\stone \mid \conone^\stone \mid (\termone^{\stone \starr \sttwo} \api \termtwo^{\stone})^{\sttwo}\\[1mm]
  & \text{\textbf{(patterns)}} & \patone,\pattwo \bnfdef {}
  & \varone^\stone \mid \conone^{\stone_1 \starr \cdots \stone_n \starr \stbase} \api \pat_1^{\stone_1} \cdots \pat_n^{\stone_n}\tspkt\\[1mm]
  & \text{\textbf{(values)}} & \valone,\valtwo \bnfdef {}
  & %
    \conone^{\stone_1 \starr \cdots \starr \stone_n \starr \stone} \api \valone_1^{\stone_1} \cdots \valone_n^{\stone_n}
    \mid\ \funone^{\stone_1 \starr \cdots \starr \stone_m \starr \stone_{m+1} \starr \stone} \api \valone_1^{\stone_1} \cdots \valone_m^{\stone_m}\tspkt
\end{align*}
% The presented operators are all standard. The pair destructor
% $\letexp{\termone}{\varone}{\vartwo}{\termtwo}$ binds the variables
% $\varone$ and $\vartwo$ to the two components of the result of
% $\termone$ in $\termtwo$. % The set of \emph{free variables}
% $\FV(\termone)$ of a term $\termone$ is defined in the usual way.  If
% $\FV(\termone) = \varnothing$, we call $\termone$ \emph{ground}.  A
% term $\termone$ is called \emph{linear}, if each variable occurs at
% most once in $\termone$.  A \emph{substitution} $\substone$ is a
% finite mapping from variables $\varone^\stone$ to terms
% $\termone^\stone$.  The substitution mapping
% $\vec{\varone}=\varone_1,\ldots,\varone_n$ to
% $\vec{\termone}=\termone_1,\ldots,\termone_n$, respectively, is
% indicated with $\substseq{\varone}{\termone}$ or
% $\substvec{\varone}{\termone}$ for short.  The variables
% $\vec{\varone}$ are called the \emph{domain} of $\substone$.  We
% denote by $\termone\substone$ the application of $\substone$ to
% $\termone$. Let-bound variables are renamed to avoid variable capture.
% \longv{The composition $\substone_2 \compose \substone_1$ of two
%   substitutions is given by the substitution that maps elements
%   $\varone$ from the domain of $\substone_1$ to
%   $(\varone\substone_1)\substone_2$.}

A \emph{program} $\progone$ over functions $\FUN$ and constructors
$\CON$ defines each function $\funone \in \FUN$ through a finite set
of \emph{equations} $l^\stone = r^\stone$, where $l$ is of the form
$\funone \api \pat_1 \api \cdots \pat_{\arity{f}}$.  We put the usual
restriction on equations that each variable occurs at most once in
$l$, i.e. that $l$ is linear, and that the variables of the
\emph{right-hand side} $r$ are all included in $l$.  To keep the
semantics short, we do not impose any order on the equations.
Instead, we require that left-hand sides defining $\funone$ are all
pairwise non-overlapping.  This ensures that our programming model is
deterministic.
We assume \emph{call-by-value} semantics.
The call-by-value reduction relation of a program $\progone$ is denoted by 
$\rew[\progone]$ and defined in the expected way, see~\cite{EV}.

Some remarks are in order before proceeding.  As standard in
functional programming, only values of base type can be destructed by
pattern matching. In a pattern, a constructor always needs to be fully
applied.  We
excluded $\lambda$-abstractions from our language. In our setting,
abstractions would only complicate the presentation without improving
on expressivity.  They can always be lifted to the top-level. Similarly,
conditionals and case-expressions would not improve upon
expressivity.

%%% Local Variables:
%%% mode: latex
%%% TeX-master: "paper"
%%% End:

%%%%%%%%%%%%%%%%%%%%%%%%%%%%%%%%%%%%%%%%%
\section{Sized Types and Their Soundness}\label{sect:STS}
%%%%%%%%%%%%%%%%%%%%%%%%%%%%%%%%%%%%%%%%%
\longv{
\newcommand{\BN}[1]{\sizeannotate{\tycon{Int}}{#1}}
\newcommand{\BL}[1]{\sizeannotate{\tycon{IntList}}{#1}}
}
This section is devoted to introducing the main object of study of
this paper, namely a sized type system for the applicative programs that we
introduced in Section~\ref{sect:APST}. We have tried to keep the
presentation of the relatively involved underlying concepts
as simple as possible.
%%%%%%%%%%%%%%%%%%%%%
\paragraph{Indices.}
%%%%%%%%%%%%%%%%%%%%%
As a first step, we make the notion of \emph{size index}, 
with which we will later annotate data types, precise. 
Let $\IS$ denote a set of first-order function symbols, the
\emph{index symbols}.  Any symbol $\ifunone \in \IS$ is associated
with a natural number $\arity{\ifunone}$, its \emph{arity}.  The set
of \emph{index terms} is generated over a countable infinite set of
\emph{index variables} $\ivarone \in \IVARS$ and index symbols
$\ifunone \in \IS$.
\begin{align*}
  & \text{\textbf{(index terms)}} & \itermone,\itermtwo \bnfdef {} \ivarone \mid \ifunone(\seq[\arity{\ifunone}]{\itermone}) \tpkt
\end{align*}
We denote by $\Var{\itermone} \subset \IVARS$ the set of variables
occurring in $\itermone$.  Substitutions mapping index variables to
index terms are called \emph{index substitutions}.  With $\isubstone$
we always denote an index substitution.  

Throughout this section, $\IS$ is kept fixed. Meaning is given to
index terms through an \emph{interpretation} $\iinter$, that maps
every $k$-ary $\ifunone \in \IS$ to a (total) and \emph{weakly
  monotonic} function $\interpretation[\iinter]{\ifunone} \ofdom
\N^{\arity{\ifunone}} \to \N$.
We suppose that $\IS$ always contains a constant
$\izero$, a unary symbol $\isucc$, and a binary symbol $+$ which we
write in infix notation below. These are always interpreted as zero,
the successor function and addition, respectively.  Our index language
encompasses the one of Hughes~et.~al~\cite{HPS:POPL:96}, where
linear expressions over natural numbers are considered.  
The interpretation of an index term $\itermone$, under an
\emph{assignment} $\assignone \ofdom \IVARS \to \N$ and an
interpretation $\iinter$, is defined recursively in the usual way:
\longshortv{%
\[
   \interpret[\iinter][\assignone]{\itermone} \defsym 
   \begin{cases}
     \assignone(\itermone) & \text{if $\itermone \in \IVARS$,}\\
     \interpretation[\iinter]{\ifunone}(\interpret[\iinter][\assignone]{\itermone_1},\dots,\interpret[\iinter][\assignone]{\itermone_k})
     & \text{if $\itermone = \ifunone(\seq[k]{\itermone})$.}
   \end{cases}
\]}{%
$\interpret[\iinter][\assignone]{\ivarone} \defsym \assignone(\ivarone)$
and 
$\interpret[\iinter][\assignone]{\ifunone(\seq[k]{\itermone})} 
\defsym \interpretation[\iinter]{\ifunone}(\interpret[\iinter][\assignone]{\itermone_1},\dots,\interpret[\iinter][\assignone]{\itermone_k})$.
}
We define $\itermone \leqs[\iinter] \itermtwo$ if
$\interpret[\iinter][\assignone]{\itermone} \leq
\interpret[\iinter][\assignone]{\itermtwo}$ holds \emph{for all}
assignments $\assignone$.
%%%%%%%%%%%%%%%%%%%%%%%%%%%%%%%%%%%%%%%%%%%%%%%%%%%%
\paragraph{Sized Types Subtyping and Type Checking.}
%%%%%%%%%%%%%%%%%%%%%%%%%%%%%%%%%%%%%%%%%%%%%%%%%%%%
The set of \emph{sized types} is given by annotating occurrences of base types 
in simple types with index terms $\itermone$, possibly introducing quantification over 
index variables.
More precise, the sets of \emph{(sized) monotypes}, \emph{(sized) polytypes} and \emph{(sized) types}
are generated from base types $\stbase$, index variables $\vec{\ivarone}$ 
and index terms $\itermone$ as follows:
\begin{align*}
  & \text{\textbf{(monotypes)}} & \mtpone,\mtptwo \bnfdef {} 
  & \base[\itermone] \mid \tpair{\mtpone}{\mtptwo} \mid \tpone \sarr \mtpone \tkom
  && \text{\textbf{(polytypes)}} & \ptpone \bnfdef {} 
  & \forall{\vec{\ivarone}}.\ \tpone \starr \mtpone \tkom
  && \text{\textbf{(types)}} & \tpone \bnfdef {} 
  & \mtpone \mid \ptpone \tpkt
\end{align*}
Monotypes $\base[\itermone]$ are called \emph{indexed base types}.
We keep the convention that the arrow binds stronger than quantification. 
Thus in a polytype $\forall{\vec{\ivarone}}.\ \tpone \starr \mtpone$ the variables $\vec{\ivarone}$ are bound in $\tpone$ and $\mtpone$. 
We will sometimes write a monotype $\mtpone$ as $\forall \seqempty.\ \mtpone$. This way, 
every type $\tpone$ can given in the form $\forall \vec{\ivarone}.\ \mtpone$. 
The \emph{skeleton} of a type $\tpone$ is the simple type obtained by dropping quantifiers and
indices.  The sets $\FPV(\cdot)$ and $\FNV(\cdot)$, of free variables
occurring in \emph{positive} and \emph{negative} positions,
respectively, are defined in the natural way.
The set of free variables in $\tpone$ is denoted by $\FV(\tpone)$.
We consider types equal up to $\alpha$-equivalence. Index substitutions are extended to sized types
in the obvious way, using $\alpha$-conversion to avoid variable capture.

We denote by $\tpone \instantiates \mtpone$ that the monotype $\mtpone$ is obtained by \emph{instantiating} 
the variables quantified in $\tpone$ with arbitrary index terms, i.e.
$\mtpone = \mtpone'\substvec{\ivarone}{\itermone}$
for some monotype $\mtpone'$ and index terms $\vec{\itermone}$, 
where $\tpone = \forall{\vec{\ivarone}}.\mtpone'$.
Notice that by our convention $\mtpone=\forall\seqempty.\ \mtpone$, 
we have $\mtpone \instantiates \mtpone$ for every monotype $\mtpone$.

\newcommand{\checkstbase}{\rl{\ensuremath{\subtypeof[{\stbase}]}}}
\newcommand{\checkstpair}{\rl{\ensuremath{\subtypeof[\times]}}}
\newcommand{\checkstarr}{\rl{\ensuremath{\subtypeof[\starr]}}}
\newcommand{\checkstforall}{\rl{\ensuremath{\subtypeof[\forall]}}}
\newcommand{\checkvarsd}{\rl{Var}}
\newcommand{\checkfunsd}{\rl{Fun}}
\newcommand{\checkappsd}{\rl{App}}
\newcommand{\checkletsd}{\rl{LetPair}}
\newcommand{\checkpairsd}{\rl{Pair}}

\begin{figure}[t]
  \begin{subfigure}[b]{1.0\linewidth}
    \centering
    \begin{framed}
      \[
        \Infer[\checkstbase]
        {\base[\itermone] \subtypeof \base[\itermtwo]}
        {\itermone \leqs \itermtwo} 
        \qquad
        \Infer[\checkstarr]
        {\tpone_1 \sarr \mtpone_1 \subtypeof \tpone_2 \sarr \mtpone_2}
        {\tpone_2 \subtypeof \tpone_1 & \mtpone_1 \subtypeof \mtpone_2}
        \qquad
        \Infer[\checkstforall]
        {\forall\vec{\ivarone}.\mtpone_1 \subtypeof \tpone_2} 
        {
          \tpone_2 \instantiates \mtpone_2 
          & \mtpone_1 \subtypeof \mtpone_2
          & \vec{\ivarone} \not\in\FV(\tpone_2)
        } 
      \]
    \end{framed}
    \caption{Subtyping rules.}\label{fig:typecheck:subtype}
  \end{subfigure}
  \\[5mm]
  \begin{subfigure}[b]{1.0\linewidth}
    \centering
    \begin{framed}
      \[
      \Infer[\checkvarsd]
            {\typedsd{\ctxone,\varone \oftype \tpone}{\varone}{\mtpone}}
            {\tpone \instantiates \mtpone} 
      \qquad\qquad\qquad
      \Infer[\checkfunsd]
            {\typedsd{\ctxone}{s}{\mtpone}} 
            {s \in \FUN \cup \CON & s \decl \tpone & \tpone \instantiates \mtpone} 
      \]
      \[
        \Infer[\checkappsd]
        {\typedsd{\ctxone}{\termone \api \termtwo}{\mtpone}}
        {
          \typedsd{\ctxone}{\termone}{(\forall \vec{\ivarone}. \mtptwo_1) \sarr \mtpone} 
          & \typedsd{\ctxone}{\termtwo}{\mtptwo_2}
          & \mtptwo_2 \subtypeof \mtptwo_1
          & \vec{\ivarone} \not\in\FV(\restrictctx{\ctxone}{\FV(\termtwo)})}
      \]
  \end{framed}
    \caption{Typing rules.}\label{fig:typecheck:type}
  \end{subfigure}
  \\
  \caption{Typing and subtyping rules, depending on the semantic interpretation $\iinter$.}\label{fig:typecheck}
\end{figure}

The subtyping relation $\subtypeof$ is given in \Cref{fig:typecheck:subtype}. 
It depends on the interpretation of size indices, but otherwise is defined in the expected way.
We are interested in certain linear types, namely those in which any
index term occurring in negative position is in fact a fresh index
variable.
\begin{definition}[Canonical Sized Type, Sized Type Declaration]\label{d:canonical}\envskipline
  \begin{varenumerate}
    \item 
      A monotype $\mtpone$ is \emph{canonical} if one of the following alternatives hold:
      \begin{varitemize}
        \item $\mtpone = \base[\itermone]$ is an indexed base type;
        % \item $\mtpone = \tpair{\mtpone_1}{\mtpone_2}$ for two canonical monotypes $\mtpone_1,\mtpone_2$; 
        \item $\mtpone = \base[\ivarone] \sarr \mtpone'$ with $\ivarone \not\in \FNV(\mtpone')$;
        \item $\mtpone = \ptpone \sarr \mtpone'$ for a canonical polytype $\ptpone$ and canonical type
          $\mtpone'$ satisfying $\FV(\ptpone) \cap \FNV(\mtpone') = \varnothing$.  
        \end{varitemize}
      \item 
      A polytype $\ptpone = \forall\vec{\ivarone}.\mtpone$ is \emph{canonical} if
      $\mtpone$ is canonical and $\FNV(\mtpone) \subseteq \{\vec{\ivarone}\}$. 
    \item       
      To each function symbol $s \in \FUN \cup \CON$, we associate a \emph{closed} and \emph{canonical} 
      type 
      $\tpone$ whose skeleton coincides with the simple type of $s$.
      We write $s \decl \tpone$ and call $s \decl \tpone$ the \emph{sized type declaration} of $s$.
    \end{varenumerate}
\end{definition}
Canonicity ensures that pattern matching can be resolved with a simple
substitution mechanism, rather than a sophisticated unification based
mechanism that takes the semantic interpretation $\iinter$ into
account. 

In \Cref{fig:typecheck:type} we depict the typing rules of our sized type system. 
A \emph{(typing) context} $\ctxone$ is a mapping 
from variables $\varone$ to types $\tpone$ so that the skeleton of $\tpone$ coincides with the simple type of $\varone$.
We denote the context $\ctxone$ that maps variables $\varone_i$ to $\tpone_i$ ($1 \leq i \leq n$)
by $\ctxseq{\varone}{\tpone}$. The empty context is denoted by $\emptyctx$.
We lift set operations as well as the notion of (positive, negative) free 
variables and application of index substitutions to contexts in the obvious way.
We denote by $\restrictctx{\ctxone}{X}$ the \emph{restriction} 
of context $\ctxone$ to a set of variables $X \subseteq \VAR$. 
The typing statement $\typedsd[\iinter]{\ctxone}{\termone}{\mtpone}$ states that under
the typing contexts $\ctxone$, the term $\termone$ has the \emph{monotype} $\mtpone$, 
when indices are interpreted with respect to $\iinter$.
%MA: added
The typing rules from \Cref{fig:typecheck:type} are fairly standard.
Symbols $s \in \FUN \cup \CON \cup \VAR$ are given instance types of their associated types. 
This way we achieve the desired degree of polymorphism outlined in Section~\ref{sect:ERW}.
Subtyping and generalisation is confined to function application, see rule~\checkappsd. 
Here, the monotype $\mtptwo_2$ of the argument term $\termtwo$ is weakened to $\mtptwo_1$, 
the side-conditions put on index variables $\vec{\ivarone}$ allow then a generalisation of $\mtptwo_1$ to $\forall\vec{\ivarone}. \mtptwo_1$,
the type expected by the function $\termone$.
This way, the complete system becomes syntax directed. 
%MA: added
We remark that subtyping is prohibited in the typing of the left spine of applicative terms.
\newcommand{\fpfun}{\rl{FpFun}}
\newcommand{\fpappvar}{\rl{FpAppVar}}
\newcommand{\fpappnvar}{\rl{FpAppNVar}}
\begin{figure}[t]
  \centering
  \begin{framed}
    \[
        \Infer[\fpfun]
          {\fpInfer{\emptyctx}{\funone}{\mtpone}}
          {\funone \decl \forall \vec{\ivarone}.\mtpone}
    \qquad\qquad\qquad\qquad 
        \Infer[\fpappvar]
          {\fpInfer{\ctxone \uplus \{\varone \oftype \tpone\}}{\termtwo \api \varone}{\mtpone}}
          {\fpInfer{\ctxone}{\termtwo}{\tpone \sarr \mtpone}}
      \]
      \[
      \Infer[\fpappnvar]
          {\fpInfer{\ctxone_1 \uplus \ctxone_2}{\termone \api \termtwo}{\mtpone\subst{\ivarone}{\itermone}}}
          {
            \begin{array}{c}
              (\FV(\ctxone_1) \cup \FV(\mtpone)) \cap (\FV(\ctxone_2) \cup \FV(\base[\itermone])) = \emptyset\\
              \fpInfer{\ctxone_1}{\termone}{\base[\ivarone] \sarr \mtpone} \qquad \fpInfer{\ctxone_2}{\termtwo}{\base[\itermone]}  \qquad \termone \not\in\VAR
            \end{array}
          }
       \]
  \end{framed}
  \caption{Rules for computing the footprint of a term.}
  \label{fig:footprint}
\end{figure}

Since our programs are equationally-defined, we need to define when
equations are well-typed. In essence, we will say that a
program $\progone$ is \emph{well-typed}, if, for all equations $l = r$, 
the right-hand side $r$ can be given a subtype of $l$.  Due to
polymorphic typing of recursion, and since our typing relation
integrates subtyping, we have to be careful.  Instead of giving $l$ an
arbitrary derivable type, we will have to give it a
\emph{most general type} that has not been weakened through subtyping. 
Put otherwise, the type for the equation, which is determined by $l$, should 
precisely relate to the declared type of the considered function.
To this end, we introduce the restricted
typing relation, the \emph{footprint relation}, depicted in
Figure~\ref{fig:footprint}. 
We are now able to state the well-typedness condition.
\begin{definition}
  Let $\progone$ be a program, such that every function and constructor has a declared sized type. 
  We call a rule $l = r$ from $\progone$ \emph{well-typed under the interpretation $\iinter$} if
  \[
    \fpInfer{\ctxone}{l}{\mtpone} \IImp \typedsd{\ctxone}{r}{\mtptwo} \textit{ for some monotype $\mtptwo$ with $\mtptwo \subtypeof \mtpone$,}
  \]
  holds for all contexts $\ctxone$ and types $\mtpone$.
  The program $\progone$ is \emph{well-typed under the interpretation $\iinter$} if 
  all its equations are.
\end{definition}

The following then gives our central result.
\begin{theorem}[Subject Reduction]\label{t:subred}
  Suppose $\progone$ is well-typed under $\iinter$.
  If $\gtyped{\termone}{\mtpone}$ and $\termone \rew[\progone] \termtwo$ then 
  $\gtyped{\termtwo}{\mtpone}$.
\end{theorem}

But what does Subject Reduction tells us, besides guaranteeing that types
are preserved along reduction? Actually, a lot: If 
$\gtyped{\termone}{\base[\itermone]}$, we are now sure that the evaluation
of $\termone$, if it terminates, would lead to a value of size at most
$\interpretation[\iinter]{\itermone}$. 
Of course, this requires that we give (first-order) \emph{data-constructors} a suitable sized type. 
To this end, let us call a sized type \emph{additive} if it is of the form
$\forall \vec{\ivarone}.\ \base[\ivarone_1] \starr \cdots \starr \base[\ivarone_k] \starr \base[\isucc(\ivarone_1 + \dots + \ivarone_k)]$.
\begin{corollary}\label{cor:size}
  Suppose $\progone$ is well-typed under the interpretation $\iinter$, 
  where data-constructors are given an additive type. 
  Suppose the first-order function $\fun{main}$ has type 
  $\forall{\vec{\ivarone}}. \base[\ivarone_1] \starr \cdots \starr \base[\ivarone_k] \starr \base[\itermone]$. 
  Then for all inputs $\seq{\dataone}$, if
  $\fun{main} \api \aseq[k]{\dataone}$ reduces to a data value $\dataone$,
  then the size of $\dataone$ is bounded by $s(\size{\dataone_1},\dots,\size{\dataone_k})$, 
  where $s$ is the function 
  $s(\seq[k]{\ivarone}) = \interpret[\iinter]{\itermone}$.
\end{corollary}
%MA: added
As we have done in the preceding examples, the notion of additive sized type could be suited so that 
constants like the list constructor $\cnil$ are attributed with a size of zero.
Thereby, the sized type for lists would reflect the length of lists. Although we take this more natural 
size measure into account in our implementation, for the sake of brevity we refrained from doing so here.
Note that the corollary by itself, does not mean much about the
\emph{complexity} of evaluating $\termone$. Through the aforementioned ticking transformation however 
allows us to reduce runtime analysis of $\progone$ to a size analysis, which in turn can be carried out with our sized type system.

%%% Local Variables:
%%% mode: latex
%%% TeX-master: "paper"
%%% End:

%%%%%%%%%%%%%%%%%%%%%
\section{Conclusions}\label{sect:C}
%%%%%%%%%%%%%%%%%%%%%
We have described a new system of sized types whose key features are
an abstract index language, and higher-rank index polymorphism.  This
allows for some more flexibility compared to similar type systems from
the literature. The introduced type system is proved to enjoy a form
of type soundness. 
We have also implemented sized type inference in a prototype tool, 
called \tool{HoSA}.
Noteworthy, inference is fully automated, i.e., does not require any form of manual size annotations.

One key motivation behind this work is achieving a form of modular
complexity analysis without sacrificing its expressive power.  This is
achieved by the adoption of a type system, which is modular and
composable by definition. Noteworthy, modularity
carries to some extend through to constraint solving.  The SCCs in the
generated constraint problem are in correspondence with the SCC of the
call-graph in the input program, and are analysed independently.

Future work definitively includes refinements to our underlying constraint solver.
It would also be interesting to see how our overall
methodology applies to different resource measures like heap size
etc. Concerning heap size analysis, this is possible by ticking
constructor allocations. It could also be worthwhile to integrate a form of
amortisation in our system. This however is left for future work.

%%% Local Variables:
%%% mode: latex
%%% TeX-master: "paper"
%%% End:

%%%%%%%%%%%%%%%%%%%%%
\bibliographystyle{eptcs}
\bibliography{references}

\end{document}